\documentclass{mn2e}
\usepackage{epsfig}
\usepackage{amsmath}
 
\title[Emitting material in $\epsilon$~Eri]{On the filling factor of emitting material in the upper atmosphere of {\boldmath $\epsilon$} Eri (K2 V)}
\author[Sim \& Jordan]{S. A. Sim\thanks{s.sim@imperial.ac.uk}$^{1,2}$ \&
C. Jordan$^2$\\
$^1$Astrophysics Group, Imperial College London,
Blackett Laboratory, Prince Consort Road, London, SW7 2BW, UK\\
$^2$Department of Physics (Theoretical Physics), University of Oxford, 1 
Keble Road, Oxford, OX1 3NP, UK}

\date{\today}

\pagerange{\pageref{firstpage}--\pageref{lastpage}}
\pubyear{2003}

\begin{document}
\maketitle
\label{firstpage}

\begin{abstract}
The emission measure distribution in the upper transition region and
corona of $\epsilon$~Eri is derived from observed emission line fluxes.
Theoretical emission measure distributions are calculated assuming that
the radiation losses are balanced by the net conductive flux. We discuss
how the area factor of the emitting regions as a function of
temperature can be derived from a comparison
between these emission measure distributions. It is found that the
filling factor varies from $\sim$ 0.2 in the mid transition region
to $\sim$ 1.0 in the inner corona. The sensitivity of these results to the
adopted ion fractions, the iron abundance and other parameters is
discussed. The area factors found are qualitatively
similar to the observed structure of the solar atmosphere, and can be used
to constrain two-component models of the chromosphere. Given further
observations, the method could be applied to investigate the trends in
filling factors with indicators of stellar activity.
\end{abstract}


\section{Introduction}

It is well known from direct observations that the solar chromosphere
and transition region are far from homogeneous (see, for example, the uv
images shown by Reeves 1976 and Gallagher et al. 1998). 
In spatially resolved observations of the Sun,
transition region
line emission 
tends to be dominated by contributions from the network boundaries.
The solar-stellar analogy tells us that a similar
effect must be expected on the
surfaces of other main-sequence stars. Therefore, it is of
interest to determine the filling factor of emitting material in 
solar-like stars and look for correlations with stellar parameters, 
such as has been done for surface (photospheric) magnetic
filling factors by Montesinos \& Jordan (1993).
Such correlations may help to identify the physical processes which control
coronal/chromospheric activity.

The current generation of high resolution spectrographic instruments
are providing spectra of unprecedented quality in the ultraviolet (uv) and
X-ray wavelength ranges. 
Unfortunately, such data cannot provide direct information about the spatial
distribution of emitting material in main-sequence
star atmospheres since it is beyond the scope of current instruments
to resolve dwarf star surfaces, with the exception of the Sun. 
There is little prospect of overcoming
this limitation in the near future.

In the absence of
spatial information, semi-empirical modelling of main-sequence star
spectra is
effectively limited to the creation of ``mean'' 1-D atmospheric models whose
parameters are those of a hypothetical homogeneous atmosphere which reproduces
the observed spectrum (e.g. the $\epsilon$~Eri models constructed by
Kelch 1978 or
Sim 2002). Although such models are useful, tighter constraints on the true
atmospheric parameters, and therefore the nature of the physical processes 
which operate, could be achieved if the distribution of emitting
material were known. (Information on the structure within the 
atmospheres of cool {\it giant} stars has been obtained through 
studies of molecular fluorescence, see McMurry \& Jordan 2000.)
The approach which could be taken is illustrated by the work of Cuntz et
al. (1999). They have explored two-component chromospheres of K2~V stars,
in which one component is heated by longitudinal tube waves and the other
by acoustic waves. They simulate the supergranulation network using a flux
tube model based on empirical correlations between observed surface
magnetic fluxes and stellar rotation rates (Cuntz, Ulmschneider \&
Musielak, 1998) and calculate the flux in the Ca~{\sc ii} 
H and K lines expected
from magnetoacoustic and acoustic waves. In the absence of observational
constraints they use an average chromospheric network filling factor of
0.305. Thus a method which gives the filling factor in the transition
region between the chromosphere and corona would provide valuable
constraints on the above type of theoretical modelling.

Kopp (1972) discussed the energy balance in the solar transition region,
in the light of early observations of the solar supergranulation
structure. He introduced an area filling factor and by comparing
emission measure distributions based on spatially
integrated observations with the predictions of the energy balance
equation he proposed that the transition region emission is
restricted to an area of about 20~per cent of the solar surface area.
In this case, the conductive flux in the mid-transition region is reduced
and the lower transition region requires an additional source of heating.
This type of structure was also explored by Gabriel (1976).
The influence of the cross-section area and magnetic field expansion
factor has also been recognized in the context of loop models of solar and
stellar active regions (see Underwood, Antiochos \& Vesecky 1981).

A method of deriving area factors has also been discussed by Jordan
(2000), in which the total pressure is not kept constant but is allowed to
vary according to hydrostatic equilibrium.
Although Philippides (1996) used this approach to estimate the filling
factor at $T_e \simeq 10^5$~K in several main-sequence stars, the
observational data available were not sufficient to find the variation of
the filling factor with $T_e$. Here we apply the method to spatially
unresolved stellar observations to derive, for the first time, the
area filling factor as a function of $T_e$ in the upper transition
region of $\epsilon$~Eri.

The star selected for this analysis, $\epsilon$~Eri (K2V), is a
nearby dwarf that has moderately high levels of chromospheric/coronal
activity and for which high quality 
spectroscopic data are available. This paper forms part of our study of the
outer atmosphere of $\epsilon$~Eri: see Jordan et al. (2001a), Jordan et al.
(2001b), Sim \& Jordan (2003) and Sim (2003, in preparation) for details.

In the next section the theory required for the analysis is discussed.
Section~3 discusses the observations that are used to constrain the emission
measure distribution (Section~4). Section~5 presents ``first-cut'' area
filling factors deduced from the data and discusses possible sources of error.
In Section~6, a self-consistent determination of the filling factors
is presented.
The results are discussed in Section~7 and conclusions are drawn in Section~8.

\section{Theory}

Spatially unresolved stellar observations can be used to constrain the 
apparent mean 
emission measure (see Pan \& Jordan 1995) distribution defined by

\begin{equation}
Em^{0.3}(T_{e}) = \int_{{T_{e}/\sqrt{2}}}^{\sqrt{2}T_{e}} N_{e}(T) 
N_{H}(T) 
\frac{\mbox{d}r}{\mbox{d}T} G(r) f(r) \frac{A(r)}{A_{*}(r)} \; \mbox{d}T
\end{equation}
where $T_{e}$ is the electron temperature at radius $r$, $N_{e}$ and $N_{H}$
are the electron and hydrogen number densities respectively, 
$f(r)=r^{2}/R_{*}^2$ ($R_{*}$ is the photospheric stellar radius and
$f(r) = 1$ at the stellar surface). 
The dilution factor $G(r)$ specifies the
fraction of photons (emitted in an optically thin line at radial coordinate 
$r$) that escape the star uninterrupted by the stellar surface. 
For spherical geometry,

\begin{equation}
G(r)= 0.5 \left( {1- \sqrt{ 1- [f(r)]^{-1}}} \right) \; \; .
\end{equation}
The $0.3$ included in the notation for the emission measure in 
equation (1) is to remind
the reader that the emission measure adopted 
is an integral over a logarithmic
temperature range $\Delta \log T_{e} = 0.3$~dex. This temperature range is
chosen since it is the typical temperature range across
which individual spectral lines form in the transition regions of cool stars.
The area filling factor $A(r)/A_{*}(r)$ specifies the fraction of the total
stellar surface area at radius $r$ ($A_{*}(r) = 4 \pi r^2$) which is occupied
by emitting material. The purpose of this paper is to show how values for
$A(r)/A_{*}(r)$ can be extracted from spatially unresolved data.

In the transition region where the pressure varies slowly as a function
of the temperature, equation~(1) can be written approximately as

\begin{equation}
Em^{0.3}(T_{e}) =G(r) f(r) \frac{A(r)}{A_{*}(r)} \frac{\mbox{d}r}{\mbox{d}
T_{e}} \frac{P_{e}P_{H}}{\sqrt{2}k_{B}^2T_{e}}
\end{equation}
where $P_{e} = N_{e} k_{B} T_{e}$ and $P_{H} = N_{H} k_{B} T_{e}$ are
the electron and hydrogen pressures 
respectively, and are approximately constant over the region where an
individual line is formed. The approximation that $\mbox{d}r/\mbox{d}T_{e}$
is constant in the region of line formation is less good, but is improved
upon when finding final models by iteration.

As discussed by Jordan (2000), it is possible to explain the complete
radiation losses from the upper transition region in terms of the
divergence of the net conductive energy flux from the overlying corona.
The justification for this is that the emission lines are formed over a
pressure-squared scale height $\sim 40,000$~km, whereas the main heating of 
the quiet corona occurs at much greater heights.
It cannot be ruled out that there is an additional source of heating 
in this part of the atmosphere, but it is certain that any additional heating
must be small. The conductive flux can be expressed as

\begin{equation}
F_{c} = - \kappa T_{e}^{5/2} \frac{\mbox{d}T_{e}}{\mbox{d}r}
\end{equation}
where $\kappa \approx 1.1 \times 10^{-6}$~ergs~cm$^{-2}$~s$^{-1}$~K$^{-7/2}$
at $T_{e} \sim 10^6$~K (Spitzer 1956). $\kappa$ 
increases at lower temperature (by about $\sim 30$ per cent by $2 \times 10^5$~K), but it is taken as constant here for simplicity. Building on the simpler
plane parallel approximation used by Jordan \& Brown (1981),
the logarithm of equation~(3) can be differentiated and the temperature
gradient replaced by the conductive flux using equation~(4). By assuming
an energy balance between conduction and radiation, the net conductive
flux can be replaced by the radiation losses and the following equation
obtained:

\begin{multline}
\frac{\mbox{d}}{\mbox{d} \log T_{e}} \log \left[ {\frac{A_{*}(r) Em^{0.3}}{A(r)}
} \right] =
\frac{\mbox{d} \log G(r)}{\mbox{d} \log T_{e}}
+ 2\frac{\mbox{d} \log f(r)}{\mbox{d} \log T_{e}} \\
+ \frac{\mbox{d} \log A(r)/A_{*}(r)}{\mbox{d} \log T_{e}} 
+ 2\frac{\mbox{d} \log P_{g}}{\mbox{d} \log T_{e}} 
+ \frac{3}{2} \\
-
\frac{2 P_{\mbox{\scriptsize rad}}(T_{e})}{\kappa P_{e} P_{H} T_{e}^{3/2} G(r)^2 f(r)^2} \left[ {\frac{A_{*}(r) Em^{0.3}}{A(r)}} \right]^2\; .
\end{multline}
$P_{\mbox{\scriptsize rad}}(T_{e})$ is the radiative power loss function 
which describes the emissivity of a plasma of
fixed composition at temperature $T_{e}$. $P_{g} = N_{g} k_{B} T_{e}$
is the total gas pressure.

This equation can be solved numerically to obtain the distribution
of $A_{*}(r) Em^{0.3}/A(r)$ as a function of $T_{e}$. By comparison
with values of $Em^{0.3}(T_{e})$ (deduced from observations) this
allows the area
filling factor $A(r)/A_{*}(r)$ to be derived.

\section{Observations}

The mean emission measure distribution can be constrained using observed
emission line fluxes. The details of this procedure are well known and have
been discussed previously by various authors 
(e.g. Jordan \& Brown 1981, Jordan et al. 1987; Griffiths \& Jordan 1998).

In this paper, the mean emission measure distribution in the upper 
transition region of $\epsilon$~Eri is
constrained using the observed fluxes of emission lines of iron. 
For the most part,
the observed fluxes used are from the analysis performed by 
Schmitt et al. (1996) using data recorded by the {\it Extreme Ultraviolet 
Explorer} ({\it EUVE}) satellite. 
The data and line flux measurements are 
discussed by Schmitt et al. (1996) and Laming, Drake \& Widing (1996).
Schmitt et al. (1996) give fluxes for Fe lines from ionisation stages
{\sc ix} -- {\sc xvi} and {\sc xviii} -- {\sc xxi}. 
For the current work, only one line from each 
ionisation stage detected in the {\it EUVE} data is used. 
For Fe~{\sc xii} the three lines
at around 193~\AA~ have been added to give the total flux in the
$^4$S --$^4$P multiplet.
In other cases where there are several
lines from the same ion reported by Schmitt et al. (1996), the strongest line
has generally been used. 
The lines used are listed in Table~1 together with the percentage
errors in the measured fluxes given by Schmitt et al. (1996).

\begin{table*}
\caption{Lines of iron observed in {\it EUVE} spectra of $\epsilon$~Eri.}
\begin{tabular}{lccc} \hline
Ion & $\lambda_{0}$~(\AA)$^a$ & Transition & \% Error$^b$\\ \hline
Fe~{\sc ix} & 171.073 & 3p$^6$ $^1$S$_{0}$ -- 3p$^5$3d $^1$P$_1$& 17 \\
Fe~{\sc x}& 174.534 & 3s$^2$3p$^5$ $^2$P$_{3/2}$ -- 3s$^2$3p$^4$($^3$P)3d $^2$D$_{5/2}$&18\\
Fe~{\sc xi}&180.408& 3s$^2$3p$^4$ $^3$P$_{2}$ --  3s$^2$3p$^3$($^4$S)3d $^3$D$_3$&19\\
Fe~{\sc xii}&192.393/193.521/195.118& 3s$^2$3p$^3$ $^4$S$_{3/2}$ -- 3s$^2$3p$^2$($^3$P)3d $^4$P$_{1/2, 3/2, 5/2}$&40\\
Fe~{\sc xiii}&203.828&3s$^2$3p$^2$ $^3$P$_{2}$ -- 3s$^2$3p3d $^3$D$_3$&38\\
Fe~{\sc xiv}&211.317&3s$^2$3p $^2$P$_{1/2}$ -- 3s$^2$3d $^2$D$_{3/2}$&18\\
Fe~{\sc xv}&284.160&3s$^2$ $^1$S$_{0}$ -- 3s3p $^1$P$_1$&7\\
Fe~{\sc xvi}&335.410&3s $^2$S$_{1/2}$ -- 3p $^2$P$_{3/2}$&11\\
Fe~{\sc xviii}&93.923& 2s$^2$2p$^5$ $^2$P$_{3/2}$ -- 2s2p$^6$ $^2$S$_{1/2}$&22\\
Fe~{\sc xix}&108.356& 2s$^2$2p$^4$ $^3$P$_{2}$ -- 2s2p$^5$ $^3$P$_2$&24\\
Fe~{\sc xx}&132.841&  2s$^2$2p$^3$ $^4$S$_{3/2}$ -- 2s2p$^4$ $^4$P$_{5/2}$&21\\
Fe~{\sc xxi}&128.752& 2s$^2$2p$^2$ $^3$P$_0$ -- 2s2p$^3$ $^3$D$_1$&40\\ \hline
\end{tabular}

\noindent $^a$ Rest wavelength.\\
\noindent $^b$The error column gives the
percentage error in the measured flux from Schmitt et al. (1996).

\end{table*}

The observed {\it EUVE} fluxes have been converted to fluxes at the stellar 
surface using the stellar parameters for $\epsilon$~Eri listed in
table~1 of Jordan et al. (2001a). Subsequent to the work carried out by 
Schmitt et al. (1996), a more accurate determination of the hydrogen
column density along the line of sight to $\epsilon$~Eri has been made by
analysis of interstellar absorption features in the Lyman~$\alpha$ line
profile (Dring et al. 1997). This gives a column density of 
$\log N(\mbox{cm}^{-2})=17.875$, which is smaller than that used by
Schmitt et al. (1996) by close to a factor of 1.5. The wavelength dependent
transmission of the interstellar medium with the new column density has
been taken from calculations by Philippides (1996) and has been used to 
correct the observed fluxes for interstellar absorption.

The {\it EUVE} data are supplemented 
by our recent detection of the forbidden ultraviolet Fe~{\sc xii} lines
in STIS spectra of $\epsilon$~Eri. These data, and the measurement of the
Fe~{\sc xii} line fluxes, have been presented elsewhere (Jordan et al. 2001a).

\section{Emission measure loci}

As discussed by Jordan \& Brown (1981),
each line flux can be used to construct a locus which places an approximate
upper bound on the mean emission measure distribution:

\begin{equation}
Em^{0.3}(T_{e}) < F_{*} / K(T_{e})
\end{equation} 
where $F_{*}$ is the stellar surface flux for the line under consideration
and $K(T_{e})$ is a contribution
function defined by

\begin{equation}
K(T_{e}) = \frac{hc}{\lambda}\frac{N_{u}}{N_{I} N_{e}}\frac{N_{I}}{N_{E}}
\frac{N_{E}}{N_{H}} A_{ul}
\end{equation}
where $\lambda$ is the wavelength of the line, $A_{ul}$ is the Einstein
coefficient for spontaneous emission in the line,
and the $N$'s are number
densities with the subscripts $u$, $I$ and $E$ 
referring to ions in the upper level in
the transition, the ionisation stage in which the transition occurs
and the 
element giving rise to the transition 
respectively.
$K(T_{e})$ has been calculated
for each of the lines listed in Table~1, and for the Fe~{\sc xii} 
1242-\AA~line. The CHIANTI
atomic database (v4; Dere et al. 1997; Young et al. 2003) was used 
to compute ${N_{u}}/{N_{I}}$ as a function of $T_{e}$.The
calculations were performed using the best value of the mean transition region
electron pressure $\log P_{e}(\mbox{cm$^{-3}$~K})= 15.68$ obtained by
Jordan et al. (2001b).
Two sets of calculations for the ionisation fractions of iron have been used
(Arnaud \& Rothenflug 1985; Arnaud \& Raymond 1992) and the stellar 
photospheric iron 
abundance
[Fe/H]=-0.09 found by Drake \& Smith (1993) has been adopted. (The solar
iron abundance was taken from Grevesse \& Sauval 1998). 
Fig.~1 shows the loci $F_{*}/K(T_{e})$ derived
from the observed fluxes using the ionisation balance calculations of
Arnaud \& Rothenflug (1985) and Fig.~2 shows the loci obtained with the
ionisation balance of Arnaud \& Raymond (1992).

\begin{figure}
\begin{center}
\epsfig{file=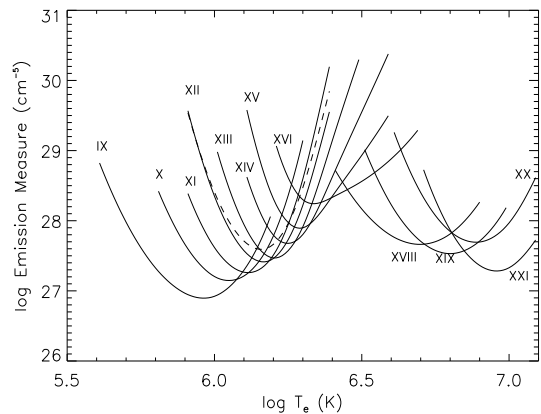, height=6cm, width=8cm}
\caption{Apparent emission measure loci deduced using the ionisation
balance of Arnaud \& Rothenflug (1985). The solid curves were computed
using the observed {\it EUVE} fluxes of Schmitt et al. (1996) and 
are labelled with the ionisation stage of iron which they represent. The
dashed curve was computed using the STIS Fe~{\sc xii} 1242-\AA~flux
reported by Jordan et al. (2001a).}
\end{center}
\end{figure}

\begin{figure}
\begin{center}
\epsfig{file=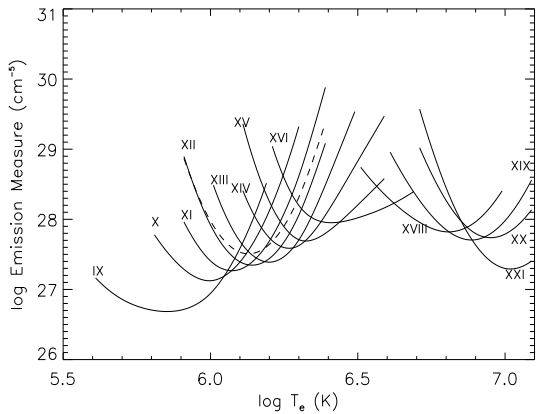, height=6cm, width=8cm}
\caption{The same as for Fig. 1, but using the ionisation
balance of Arnaud \& Raymond (1992).}
\end{center}
\end{figure}

As expected, the emission measure loci suggest a smooth
emission measure distribution
which increases with temperature up to a peak at several million 
degrees. The temperature of the peak emission measure
is difficult to determine since there are no Fe~{\sc xvii}
lines in the data which would span the critical temperatures range.
Comparing Figs.~1 and 2 shows that the choice of ionisation balance 
affects the form of the emission measure distribution.
In particular there are significant differences in the ion populations of
Fe~{\sc ix}, {\sc xvi}
and {\sc xviii} which alter the apparent shape. 
The Arnaud \& Raymond (1992) ionisation balance calculations suggest 
a higher peak
temperature ($\log T_{e}(\mbox{K})\sim 6.6$) than those of
Arnaud \& Rothenflug (1985) ($\log T_{e}(\mbox{K}) \sim 6.4$). In their 
analysis, Laming et al. (1996) adopted the Arnaud \& Rothenflug (1985)
ionisation balance (leading them to a coronal temperature 
$\log T_{e}(\mbox{K}) \sim 6.4$) since solar data suggests that these
calculations are more appropriate than those of Arnaud \& Raymond (1992) for 
Fe~{\sc ix} (see the discussion by Laming, Drake \& Widing 1995). In this
paper, all calculations are carried out
with both sets of ionisation fractions,
and the differences between the results are assumed to be
indicative of the 
errors in the 
ionisation balance.

Above the peak temperature the emission measure drops with increasing 
temperature. There appear to be
some discrepancies between the loci for the high temperature ions. 
In particular 
the Fe~{\sc xx} locus appears to lie too high compared to those of 
Fe~{\sc xix} and Fe~{\sc xxi}.
Material at temperatures which are significantly greater than that of the 
peak emission measure 
most likely occurs in stellar active regions, where it is
magnetically confined in hot loops (such as discussed by
Schmitt et al. 1996) and will not be discussed further here.
Solar observations support this interpretation.

The locus from the 1242-\AA~STIS line lies above that from the {\it EUVE}
Fe~{\sc xii} lines by about a factor of 1.5. This is significant but much
smaller than the discrepancy reported by Jordan et al. (2001a) (they 
reported a factor of 3 difference between the STIS and {\it EUVE} 
results). Jordan et al. (2001a) computed the {\it EUVE} Fe~{\sc xii}
locus using the flux reported by Laming et al. (1996) for the 195-\AA~line
only. The {\it EUVE} Fe~{\sc xii} loci shown in Figs. 1 and 2 are computed
using the fluxes for all three lines in the $^4$P -- $^4$S multiplet
from Schmitt et al. (1996). If the three fluxes reported by
Schmitt et al. (1996) are used separately, the 195-\AA~line gives a locus
that is significantly higher than that obtained by combining all three lines 
of the multiplet while the 192-\AA~line gives a locus that is significantly
lower. The 193-\AA~line gives a locus that agrees well with that of the 
complete multiplet. 
It is plausible that the 
discrepancy (factor of 1.5) between the {\it EUVE} and STIS
Fe~{\sc xii} loci is the result of systematic errors in the atomic data, but
the measurement errors for the {\it EUVE} fluxes (see 
Table~1) are large enough to account for most of the difference.
As suggested by Jordan et al. (2001a), there may be a real difference
related to variations in coronal activity between the times at which the
two sets of observations were made, but the uncertainty in the {\it EUVE} 
flux is too large to establish this with certainty.

\section{Emission measure models}

\subsection{Models}

As discussed in Section 2, equation~(5) can be used to construct 
theoretical emission measure distributions under the assumption of an
energy balance between thermal conduction and radiation in the upper
transition region. To determine a unique emission measure distribution from
equation~(5) two boundary conditions are required: the pressure and 
emission measure must both be fixed at some temperature. 
The pressure at all other points 
is then determined by enforcing hydrostatic equilibrium. 
For this,
a contribution to the pressure from turbulent motions is included, calculated
using a most-probable turbulent velocity $\xi \approx 21$~km~s$^{-1}$ which is
appropriate for the temperature range under consideration (Sim \& Jordan 2003).
(Here, this turbulent pressure contribution is very small.)
With the pressure so determined, equation~(5) can  
be integrated numerically to determine the emission measure at all
other points.

The first boundary condition adopted is
that the electron pressure at $\log T_{e}(\mbox{K})=5.3$ should be consistent
with the mean value determined by Jordan et al. (2001b). The effects of
slightly varying the choice of fixed electron pressure at 
$\log T_{e}(\mbox{K})=5.3$ have been explored (see below).

It is a natural consequence of the signs of the terms in equation~(5) that
the computed emission measure distribution will pass through a minimum in the 
transition region when the magnitude of the last term in equation~(5) is
close to $3/2$. 
Therefore, the second boundary condition has been chosen to be that
this minimum should occur at $\log T_{e}(\mbox{K}) = 5.3$, which
is approximately the temperature at which the minimum in the emission 
measure distribution is observed to occur in the Sun (see e.g.
Macpherson \& Jordan 1999).
The emission measure distribution for $\epsilon$~Eri (Sim 2002)
shows a decrease up to at least $\log T_{e}(\mbox{K}) = 5.3$.

Using these two boundary conditions, emission measure distributions 
($A_{*}(r) Em^{0.3}/A(r)$) have been computed using equation~(5).
At this stage, the third term on the right-hand side (RHS) of equation~(5) 
which accounts for the variation in the area factor with $T_{e}$
has been
ignored. The effect of this term is discussed in Section~6.
The distributions have been calculated from an upper boundary temperature
$\log T_{\mbox{\scriptsize max}} = 6.5$ down to a lower boundary
temperature $\log T_{\mbox{\scriptsize min}}=5.3$. 
Distributions of $A_{*}(r) Em^{0.3}/A(r)$ were calculated using a range
of starting values for $A_{*}(r) Em^{0.3}/A(r)$ and the electron pressure
at the top point in the model in order to find the distributions which
satisfies the chosen boundary conditions (see above).
Two of the computed distributions are shown in Fig.~3 (the solid and
dotted lines). These distributions
differ only in the choice of fixed pressure at the bottom end.
Fig.~4 shows the electron pressure ($P_{e} = N_{e} T_{e}$) as a
function of $T_{e}$ for the same models. The first model (solid line in the
figures) has a pressure which is slightly lower than, but close to,
the best pressure ($\log P_{e}(\mbox{cm$^{-3}$~K}) \sim 15.68$) while the
other (dotted line) has a higher pressure, just within the error margins 
of the best pressure ($\log P_{e}(\mbox{cm$^{-3}$~K}) \sim 15.78$).

In both models, it can be seen that above the mid-transition region the 
theoretical
emission measure increases roughly proportional to $T^{3/2}$, owing
to the dominance of the penultimate term in equation~(5). 
At higher temperatures, the emission measure begins to
flatten off and eventually starts to decrease. This is the result of the
negative contributions to the RHS of equation~(5) from the 1st and 4th
terms which become significant at high temperatures (the last term is
small at high temperatures).

\begin{figure}
\begin{center}
\epsfig{file=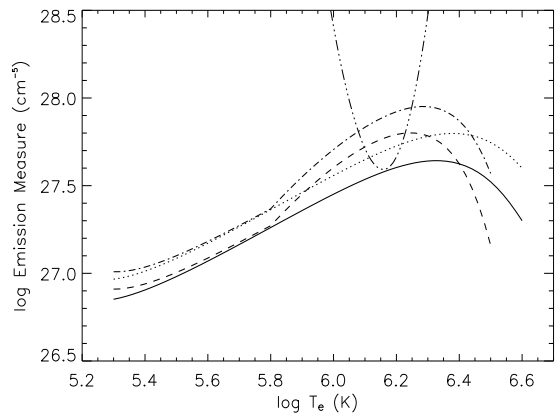, height=6cm, width=8cm}
\caption{Theoretical curves of $A_{*}(r) Em^{0.3}/A(r)$ against
temperature. The dash-triple-dot 
curve is the Fe~{\sc xii} STIS locus from
Fig.~1. The solid and dotted curves are for models in which the variation of
the filling factor with temperature (on the RHS of eqn.~(5)) has {\it not}
been included, and refer to different (higher and lower) pressures at
$\log T_{e} = 5.3$. The dashed and dashed-dot curves are for the
corresponding self-consistent models which include the variation of
the filling factor with temperature (see Section 6). }
\end{center}
\end{figure}

\begin{figure}
\begin{center}
\epsfig{file=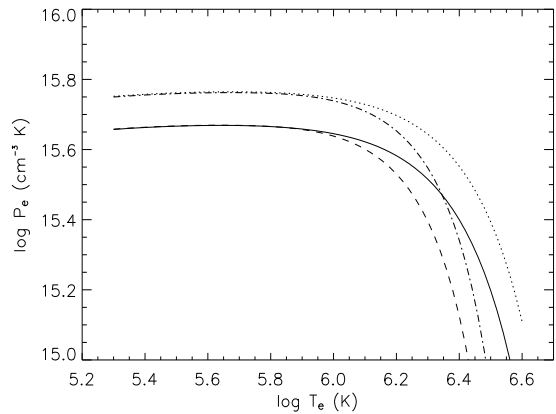, height=6cm, width=8cm}
\caption{The electron pressure as a function of the electron
temperature for the models plotted in Fig.~3, using the same symbols.}
\end{center}
\end{figure}

\subsection{Filling factors}

The theoretical distributions plotted in Fig.~3 can be used to
estimate
fractional filling factors for each of the observed iron lines. The 
apparent
emission measure distribution can be used to compute approximate line
fluxes via

\begin{equation}
F_{*} \approx \int K(T_{e}) Em^{0.3}(T_{e}) \; \mbox{d}\log T_{e}/0.3
\end{equation}
where the integral runs over the whole temperature range of the model. If 
the apparent emission measure in equation~(8) is replaced with the computed
distribution of $A_{*}(r) Em^{0.3}/A(r)$ then
$F_{*} A_{*}(r)/A(r)$
is obtained instead of $F_{*}$. 
This can be compared to the observed value of the 
line flux to deduce $A(r)/A_*(r)$ in the region of line formation. 
The value so deduced is only an estimate because of the approximate
nature of equation~(8) and must be interpreted as a
weighted average of the area filling factor over the region of line
formation. It should be a good estimate provided that the physical
extent of the region of
line formation is reasonably small (as is the case for all the lines
considered here).

The filling factors deduced from the distributions discussed above 
are given in Table~2. Filling factors are only given for 
individual lines that form
in the upper transition region (Fe~{\sc ix} -- Fe~{\sc xvi}) since the
higher temperature lines probably form under different conditions in hot 
active region loops (see Section 4). 
For each line the approximate temperature of line formation
(determined from the emission measure loci) and 
four filling factors are given. 
These correspond to the four possible combinations of the two models
and the two sets of ionisation balance calculations
(Arnaud \& Rothenflug 1985; Arnaud \& Raymond 1992).
In addition to the filling factors deduced from the upper transition region
lines, filling
factors are given in Table~2
for the mid-transition region ($\log T_{e}(\mbox{K})
\sim 5.3$) which were deduced by comparing the computed 
$A_{*}(r) Em^{0.3}/A(r)$ with values of $Em^{0.3}$ from a mean 
transition region emission measure distribution constructed from
STIS data (Sim 2002). This distribution will
be discussed in a forthcoming paper (Sim 2003, in preparation). 

\begin{table*}
\caption{Filling factors deduced from the observed $\epsilon$~Eri lines
and the first two models shown in Fig.~3. 
Columns 4--7 give the filling factors deduced from the
different combinations of models (higher/lower transition region pressure)
and ionisation balance calculations (Arnaud \& Rothenflug 1985 (AR85) and
Arnaud \& Raymond 1992 (AR92)).
The first part of the table gives the results from the lines observed
with {\it EUVE}. The second part shows the results from the Fe~{\sc xii}
1242-\AA~line (from STIS) and also from comparison with the mean 
emission measure distribution in the mid transition region (see
text). Here $\log g = 4.75$ is used (from Drake \& Smith 1993).}
\begin{tabular}{lcccccc} \hline
&&&\multicolumn{4}{|c|}{$A(r)/A_{*}(r)$}\\
Ion & $\lambda_{0}$~(\AA)$^a$ & $\log T_{\mbox{\scriptsize form}}(\mbox{K})$ & \multicolumn{2}{|c|}{$\log P_e \sim 15.68$} & \multicolumn{2}{|c|}{$\log P_e \sim 15.78$}\\
& & &AR85&AR92&AR85&AR92\\ \hline
Fe~{\sc ix} & 171.073 & 5.9&0.34&0.20&0.26&0.16\\
Fe~{\sc x}& 174.534 & 6.0&0.59&0.58&0.45&0.45\\
Fe~{\sc xi}&180.408& 6.1&0.76&0.74&0.58&0.57\\
Fe~{\sc xii}&192.393/193.521/195.118& 6.15&1.1&0.83&0.84&0.63\\
Fe~{\sc xiii}&203.828&6.2&1.3&0.89&0.96&0.66\\
Fe~{\sc xiv}&211.317&6.27&2.0&1.3&1.5&0.96\\
Fe~{\sc xv}&284.160&6.33&3.1&1.5&2.2&1.0\\
Fe~{\sc xvi}&335.410&6.4&5.4&2.4&3.6&1.5\\
\hline
\hline
Fe~{\sc xii}& 1242.00&6.15&1.7&1.2&1.3&0.92\\
\multicolumn{2}{|c|}{TR EMD}&5.3& \multicolumn{2}{|c|}{0.29}&
\multicolumn{2}{|c|}{0.22}\\ \hline
\end{tabular}

\noindent $^a$ Rest wavelength.

\end{table*}

The models suggest that  
the mid transition region filling factor is several tens per cent
and that the filling factor starts increasing with temperature
above $\log T_{e}(\mbox{K}) \sim 5.9$, becoming close to 1.0 at
coronal temperatures. 
To illustrate this point,
Fig.~5 shows (open symbols) 
a plot of the area filling factors deduced from the model
with higher pressure using the Arnaud \& Raymond (1992) ionisation balance
(the error bars indicate the random measurement errors but do not attempt to
address the potential systematic errors which are discussed below).
This trend with temperature is striking and consistent with what would be
expected from the standard model of expanding magnetic funnels in the 
upper atmosphere (Gabriel 1976).

\begin{figure}
\begin{center}
\epsfig{file=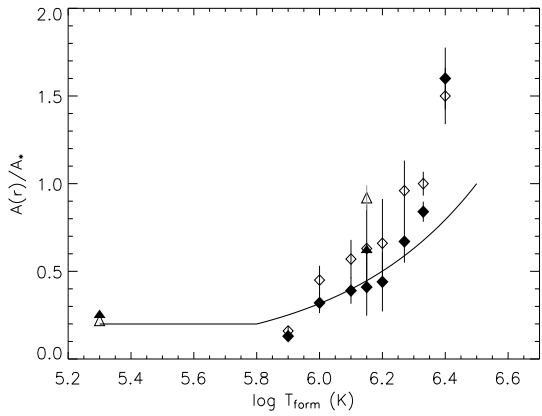, height=6cm, width=8cm}
\caption{Area filling factors deduced from the model with
$\log P_{e} \sim 15.78$ using the Arnaud \& Raymond (1992) ionisation
balance calculations. The open diamonds show the results from the {\it EUVE}
iron lines. The open triangles show the results from the STIS data
(the Fe~{\sc xii} 1242-\AA~line and the mid-transition region lines).
These results do not include the term in equation~(5) describing the 
variation of the area factor with temperature. The filled symbols
show the results when the variation of the filling factor with
temperature is included in a self-consistent manner (see Section 6).}
\end{center}
\end{figure}

The filling factors are not strongly sensitive to the adopted 
boundary pressure (they are 20 -- 40 per cent smaller at the 
larger pressure). As discussed above, however, several of the filling
factors are quite sensitive to the adopted ionisation balance. 

On physical grounds, the filling factor cannot be greater than 1.0.
In all cases the filling factor calculated for Fe~{\sc xvi} is greater
than 1.0 and, for all cases except the high pressure/Arnaud \& Raymond 
(1992) combination, the filling factor is unphysical for some of the
other lines.
In most cases the random measurement errors are not sufficient to 
explain these unphysical results, and they point towards a systematic error
in the calculations. Several possible sources of systematic errors are 
discussed below.

\subsection{Systematic errors}

\subsubsection{Ionisation balance}

The quality of the atomic data is probably the most significant source of
uncertainty, particularly those used
in the ionisation balance. By comparing the filling
factors deduced using Arnaud \& Rothenflug (1985) and Arnaud \& Raymond
(1992) it is clear that for several of the ions (Fe~{\sc ix}, Fe~{\sc xv}
and Fe~{\sc xvi}) the uncertainty in the ion balance may be comparable to,
or even greater than, a factor of 2. For Fe~{\sc xii} and Fe~{\sc xiii}
the uncertainty is smaller ($\sim 40$ per cent) and it is negligible for
Fe~{\sc x} and Fe~{\sc xi}. 

\subsubsection{Hydrogen column density}

The results are sensitive to the adopted interstellar hydrogen column
density since it determines the wavelength dependent transmission factor
of the interstellar medium. 
This is most important for Fe~{\sc xvi} (the longest wavelength
line). The error on the column density quoted by Dring et al. (1997) is
only 20 per cent, however, and so this should not be a dominant
contribution to the uncertainty. 

\subsubsection{The abundance of iron}

A potentially important source of systematic error is the adopted
iron abundance. The quoted error in the photospheric 
abundance derived by Drake \& Smith (1993) 
is only 0.05~dex, but it is possible that the iron abundance in
the upper transition region is different from that in the photosphere.
It has been suggested that the abundances of elements with low ($<10$~eV)
first ionisation potentials (FIPs)
are enhanced in the outer parts of cool star atmospheres
relative to elements with high FIPs (Crawford, Price \& Sullivan 1972 first
identified this effect in the solar wind).
This is known as the FIP-effect (see Jordan et al. 1998 and references therein
for a full discussion).
Laming, Drake \& Widing (1996) were unable to conclusively detect
an enhancement of the iron abundance in the corona of
$\epsilon$~Eri
 (as would be expected if the FIP-effect were acting),
however they could not rule out an enhancement of similar magnitude to that
in the Sun (increase by a factor of 2 -- 3). 
If the upper transition region/coronal
abundance is enhanced relative to the photospheric value adopted in the
models, then the predicted area filling factors in Table~2
will all be systematically
overestimated.
(It can be shown from equation~(5) that an increase in the adopted iron
abundance by a factor $x$ would lead to a reduction in the derived area
factors by approximately a factor of $\sqrt{x}$.)

\subsubsection{The stellar surface gravity}

Errors in the assumed value for the stellar surface gravity $g$ will affect
the calculated area filling factors in a systematic way. In particular, if
the adopted gravity is too large, the area factors found for the high
temperature lines will be too large. This is because the magnitude of the
4th term on the RHS of equation~(5) is proportional to the adopted gravity.
Since that term is negative, if the adopted gravity is too large it will mean 
that the peak in the calculated emission measure distribution will 
occur at too low a temperature (recall that the 4th term is 
important in determining where the peak occurs). To investigate this effect
models have been computed using a lower surface gravity, $\log g=4.65$,
which is consistent with the error bar on the value of the surface gravity
found by Drake \& Smith (1993) ($\log g = 4.75 \pm 0.1$). 
The emission measure distributions obtained with the lower gravity (using
the same lower boundary pressures as before) are plotted in Fig.~6.
These emission measure distributions have been used to compute area filling
factors (using the Arnaud \& Raymond 1992 ionisation balance) for comparison
with those found in Table~2. The area factors deduced for the EUVE lines
are given in Table~3.

\begin{figure}
\begin{center}
\epsfig{file=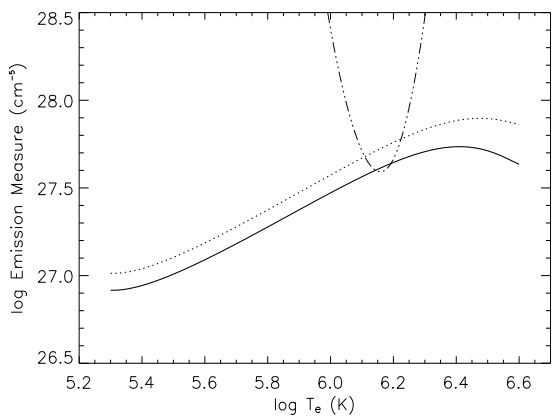, height=6cm, width=8cm}
\caption{The same as for the solid and dotted lines in
Fig. 3, but using a lower gravity of $\log g
= 4.65$.}
\end{center}
\end{figure}

\begin{table*}
\caption{Filling factors deduced from the iron lines observed in 
$\epsilon$~Eri and the models shown in Fig.~6. Columns 4 and 5 give
the filling factors deduced from the two models (higher/lower
transition region pressure) using the ionisation balance calculations
of Arnaud \& Raymond (1992) for the {\it EUVE} lines. Here $\log g =
4.65$ is used.}
\begin{tabular}{lcccc} \hline
&&&\multicolumn{2}{|c|}{$A(r)/A_{*}(r)$}\\
Ion & $\lambda_{0}$ (\AA)~$^a$ & $\log T_{\mbox{\scriptsize form}}(\mbox{K})$ & \multicolumn{1}{|c|}{$\log P_e \sim 15.68$} & \multicolumn{1}{|c|}{$\log P_e \sim 15.78$}\\
\hline
Fe~{\sc ix} & 171.073 & 5.9&0.20&0.16\\
Fe~{\sc x}& 174.534 & 6.0&0.55&0.43\\
Fe~{\sc xi}&180.408& 6.1&0.70&0.54\\
Fe~{\sc xii}&192.393/193.521/195.118& 6.15&0.76&0.59\\
Fe~{\sc xiii}&203.828&6.2&0.79&0.61\\
Fe~{\sc xiv}&211.317&6.27&1.1&0.85\\
Fe~{\sc xv}&284.160&6.33&1.2&0.86\\
Fe~{\sc xvi}&335.410&6.4&1.6&1.1\\
\hline
\end{tabular}

\noindent $^a$ Rest wavelength.

\end{table*}

Comparing the area factors in Table~3 with the appropriate entries in Table~2
shows that, as expected, the high temperature lines have the greatest 
sensitivity
to the adopted gravity: for Fe~{\sc ix} -- Fe~{\sc
xii} the difference in the adopted gravity decreases the computed area factor
by less than 10 per cent, but for Fe~{\sc xvi} the reduction is close to 
50 per cent.

Therefore, it is possible that the unphysical area factors for the high 
temperature Fe~{\sc xvi} line could arise because the adopted surface
gravity is too large. (Note that earlier work, e.g. by Kelch 1978 did adopt
the significantly lower surface gravity of $\log g=4.5$.)

\subsubsection{The radiative power loss function, $P_{\mbox{\scriptsize rad}}
(T_{e})$}

In all the calculations discussed here, 
the fit to the radiation loss function

\begin{equation}
P_{\mbox{\scriptsize rad}}(T_{e}) = 1.25 \times 10^{-16} T_{e}^{-1} \; 
\mbox{ergs~cm$^{3}$~s$^{-1}$}
\end{equation}
derived by Philippides (1996) has been adopted. 
Across the temperature range under consideration, this fit is
accurate to within a factor of $\sim 2$.
For future work, in which
it is hoped that the other sources of systematic error may be 
eliminated, it would be valuable to obtain a more accurate
power loss function using modern atomic data. For the 
present, however, this simple form is sufficiently accurate given the
uncertainties in the ionisation fractions and iron abundance, and
that the last term in equation~(5) is only important at the low temperature
end of the models.

\subsubsection{Discussion of systematic errors}

To conclude this section, it seems probable that the uncertainties in the deduced
filling factors are dominated by those in the adopted ionisation balance
calculations.
The uncertainty in the Fe~{\sc xvi} filling factor is likely to be the
largest (a multiplicative factor of 2 -- 3)
since it is the most sensitive to errors in the adopted 
hydrogen column density, surface gravity and
has amongst the most uncertain ionisation fraction. These sources of error
are certainly large enough to explain the moderately unphysical filling
factors deduced for this line.

The errors are likely to be smallest in the Fe~{\sc x}, Fe~{\sc xi},
Fe~{\sc xii} and Fe~{\sc xiii} filling factors since these are the least
sensitive to the systematic errors discussed above.

The scale of the uncertainties in all cases is small enough that 
the filling factors presented in
Table~2 can be regarded as reasonable estimates. 
That the higher pressure model with the Arnaud
\& Raymond (1992) ionisation balance gives the ``most physical'' set of filling
factors could be interpreted as meaning that this is the best combination
of pressure and ionisation balance, but the errors are large enough that this
cannot be clearly established.

The detected trend with temperature is not expected to be
the result of any of the errors discussed here and is almost certainly real: 
$A(r)/A_*(r)$ varies systematically by more than a factor of 5 over the
temperature range considered. This variation is larger than that
expected for any of the uncertainties discussed above.

\section{Self-consistent calculation}

The estimates for $A(r)/A_*(r)$ presented above are not self-consistent.
This is because when equation~(5) was used the 3rd term on the RHS was
neglected, but the implied filling factors suggest a significant variation
of $A(r)/A_*(r)$ with $\log T_{e}$. To investigate this inconsistency, 
two more models have been constructed accounting for the previously
neglected term.

To do this, the area filling factor must be specified as a function of
temperature in the modelling. In principle self-consistency between the
variation of the area factor adopted in the modelling and the derived area
factors could be achieved iteratively, but since there are significant
uncertainties (see above) a rigorous iterative solution is not warranted
and only approximate self-consistency is pursued. To this end
a simple form for $A(r)/A_*(r)$ as a function of $T_e$ has been assumed 
in the calculation of the two new models, based on the approximate
area factors presented in Table~2. Given the form of the relevant term in
equation~(5), it is computationally convenient to express the relationship
as a power law. The relationship adopted in the calculations is

\begin{equation}
\log \frac{A(r)}{A_*(r)} = \left\{ {\begin{array}{ll}
-0.7 & \log T_e < 5.8 \\
-6.5+\log T_e& 5.8 < \log T_e < 6.5 \\
\end{array} }\right. \;\; .
\end{equation}

This relationship crudely describes the form suggested by the results in
the last column of
Table~2: constant below $\log T_e = 5.8$ and then increasing to
approximately 1.0 in the corona. 
Distributions of 
$A_{*}(r) Em^{0.3}/A(r)$ calculated from the two new models (which, like the
first two models differ only in the adopted lower boundary pressure)
are plotted in Fig.~3 (the dashed and dashed-dot lines). The electron 
pressures in these models are plotted in Fig.~4.

Comparing the distributions in Fig.~3 shows the 
effect of the area variation term:
it steepens the predicted emission measure gradient before the
peak (the kink in the dashed and dashed-dot lines
is not physical but due to the discontinuity in equation~10).
Also, above the peak the emission measure declines more rapidly. This is
because the larger emission measures around the peak lead to a more
rapid decline in the
pressure with temperature, making the magnitude
of the 4th term on the RHS of equation~(5) greater. 

Table~4 gives the area factors deduced from the new models in the same format
as those given in Table~2. The area factors deduced from the new model with 
the higher pressure and the Arnaud \& Raymond (1992) ionisation balance are 
shown in Fig.~5 (filled symbols). A curve showing the assumed 
$A(r)/A_*(r)$ (equation 10) is over-plotted for comparison. 

\begin{table*}
\caption{As for Table 2, but using the self-consistent models shown in
Fig. 3.}
\begin{tabular}{lcccccc} \hline
&&&\multicolumn{4}{|c|}{$A(r)/A_{*}(r)$}\\
Ion & $\lambda_{0}$~(\AA)$^a$ & $\log T_{\mbox{\scriptsize form}}(\mbox{K})$ & \multicolumn{2}{|c|}{$\log P_e \sim 15.68$} & \multicolumn{2}{|c|}{$\log P_e \sim 15.78$}\\
& & &AR85&AR92&AR85&AR92\\ \hline
Fe~{\sc ix} & 171.073 & 5.9& 0.25&0.17&0.20&0.13\\
Fe~{\sc x}& 174.534 & 6.0&0.40&0.41&0.31&0.32\\
Fe~{\sc xi}&180.408& 6.1&0.50&0.50&0.38&0.39\\
Fe~{\sc xii}&192.393/193.521/195.118& 6.15&0.73&0.55&0.55&0.41\\
Fe~{\sc xiii}&203.828&6.2&0.86&0.60&0.63&0.44\\
Fe~{\sc xiv}&211.317&6.27&1.4&0.96&0.99&0.67\\
Fe~{\sc xv}&284.160&6.33&2.4&1.33&1.6&0.84\\
Fe~{\sc xvi}&335.410&6.4&5.2&2.9&3.2&1.6\\
\hline
\hline
Fe~{\sc xii}& 1242.00&6.15&1.1&0.81&0.82&0.61\\
\multicolumn{2}{|c|}{TR EMD}&5.3& \multicolumn{2}{|c|}{0.33}&
\multicolumn{2}{|c|}{0.24}\\ \hline
\end{tabular}

\noindent $^a$ Rest wavelength.

\end{table*}

Given that there are several sources of
systematic error, the agreement between the form of $A(r)/A_*(r)$ used in the
modelling and the values of $A(r)/A_*(r)$ predicted by the model is 
sufficiently good that the higher pressure model can be regarded as 
self-consistent if the Arnaud \& Raymond (1992) ionisation balance is 
adopted. It is noteworthy that the largest discrepancies are for
Fe~{\sc ix}, Fe~{\sc xv} and, in particular, Fe~{\sc xvi} which are the ions
that show the largest uncertainties in the ionisation balance.
The area factors deduced from the model with lower pressure are slightly less 
self-consistent than those obtained from the higher pressure model, being
systematically larger than those assumed in the modelling. Given the
scale of the other sources of error, however, the agreement is adequate.

Comparing Tables~2 and 4, it is clear that the main effect of 
including the term describing the variation of the area factor with
temperature is to reduce the implied area factors for most of the 
high temperature lines, leading to fewer unphysical
($A(r)/A_*(r) > 1.0$) values. This is the result of the steeper 
increase of emission measure with temperature.
The only exception is Fe~{\sc xvi} for which the implied 
area factor is {\it larger} when the variation of the area factor is
included in the model if the Arnaud \& Raymond (1992)
ionisation fractions are used. This is
because Fe~{\sc xvi} forms at temperatures above the peak of the 
emission measure distribution, 
if the Arnaud \& Raymond (1992) ionisation balance calculations
are adopted.

In summary, the self-consistent treatment of the variation of the area
factor presented in this section indicates that neglecting this variation
in the modelling may cause the area factors in the upper transition region
to be systematically overestimated to a small, but significant, extent.
Accounting for the variation of the area factors does not, however, alter
the apparent trend that the area factor increases across the upper transition
region. For now it is concluded that the area factors presented in Table~4
(or equivalently described by equation 10) are the best current estimates 
of the area filling factors in the upper transition region of $\epsilon$~Eri
and that most of them are accurate to within a factor of two.

\section{Discussion and prospects for future development}

In Sections 5 and 6 it has been shown that the theory described in
Section 2 can be used to provide estimates of the area filling factors of
emitting material at various temperatures in stellar transition regions.

The results show (Fig.~5) that 
in $\epsilon$~Eri
the area filling factor increases with 
temperature markedly in the upper transition region, from $\sim 0.2$ at
$\log T_e < 5.8$ to $\sim 1.0$ in the corona. There is no evidence for 
significant variations across the mid-transition region $5.3 <\log T_e <5.8$.

The variation of the area factor derived for $\epsilon$~Eri is similar to
that found from observations of the solar supergranulation network made
with the {\it Solar and Heliospheric Observatory} ({\it SOHO}) (Gallagher
et al. 1998). They find (see their figs. 5 and 8, in particular) that
at the resolution of the Coronal Diagnostic Spectrometer (CDS) ($\simeq$
2~arcsec by 2~arcsec) the area occupied by the emitting material increases
dramatically in the upper transition, but does not vary significantly
across the mid-transition region. These results are in general
agreement with earlier observations and the model by Gabriel (1976).
When the method used here is applied to CDS observations of the average
network, a filling factor of $\simeq$ 0.2 at $T_e \simeq 2 \times
10^5$~K is deduced {\it within} the network (Smith \& Jordan 2002), which
typically occupies $\simeq$ 50 per cent of the solar area at this
temperature (Gallagher et al. 1998).
The filling factors which we have derived are based on the assumption
that the observed
emission is dominated by the regions occupying the small areas. These
could refer to structures within the supergranulation network of
$\epsilon$~Eri and may not be comparable to the total solar network area
observed with the resolution of CDS. In this case, the area factors are
{\it larger} than those found for the Sun.

Using the solar analogy, the area factor found at $T_e \simeq 2 \times
10^5$~K can be applied to the lower transition region. This increases the
intrinsic radiation losses and since thermal conduction now plays
no role below $\simeq 10^5$~K, an additional source of heating is
required. As found by Sim \& Jordan (2003), {\it if} the observed
non-thermal emission line widths are associated with an Alfv\'en
wave (or fast-mode wave) flux, there is ample energy to heat both the
corona and the lower transition region, although uncertainties concerning
wave propagation and dissipation remain to be resolved.

The filling factor of $\simeq 0.2$ found for the mid-transition region of
epsilon Eri can be compared with the value of 0.3 assumed for the
chromosphere by Cuntz et al. (1999). Since the filling factor of magnetic flux
tubes should increase with height, our results suggest that a smaller
chromospheric filling factor of $\leq 0.2$ would be more appropriate; this
could be explored in future two-component models.

The accuracy of the results presented in this paper is limited by the
various systematic sources of error which were discussed in Section 5.3.
In particular
the analysis of larger sets of lines, ideally covering a wide range
of elements and ionisation species would help. This would reduce the
systematic errors associated with uncertainties in the ionisation fractions
for iron
and possibly help to identify possible errors arising from the 
adopted elemental abundance. Suitable data for such analysis are gradually
becoming available for a range of stars thanks to the current generation of
X-ray satellite missions. X-ray data are also advantageous since their 
analysis
is less sensitive to the assumed interstellar hydrogen column density.
Improved calculations of atomic data for iron, and the radiation loss 
function 
$P_{\mbox{\scriptsize rad}}$ would also make the modelling more 
quantitatively reliable.

In
future work this technique should be applied
to a range of late-type stars with differing levels of activity.
By doing so it should be possible to look for trends in the behaviour of the
filling factors of emitting material with various stellar parameters.
The ability to reproduce these trends will then be a new constraint on 
theoretical studies of the structure and heating of late-type stars, including the magnetic field geometry.

\section{Conclusions}

It has been shown that by assuming an energy balance between radiation and
thermal conduction in the upper transition region, it is possible to use
spatially unresolved stellar observations to estimate the fractional area
filling factor of emitting material in stellar atmospheres.

The technique has been applied to $\epsilon$~Eri and it has been found that
the area filling factor is $\sim 0.2$ in the mid transition region and that
it increases to $\sim 1.0$ in the corona. 
These filling factors are similar to those found from observations of 
the solar transition region, and from models in which the 
transition region consists of material in magnetic funnels whose radii
are greatest is in the upper transition region/corona. 
Several potential sources of uncertainty in
the calculations have been discussed and it is concluded that, although
they are currently significant, they do not prevent a useful application of
the methods adopted. Further observations and calculations of atomic data should be able to reduce the current uncertainties.

\section*{Acknowledgments}

This work was carried out while SAS was a research student in
the Physics Department (Theoretical Physics) at the University of Oxford,
supported by a PPARC studentship (PPA/S/S/1999/02862) and was completed  
while he was a PPARC funded PDRA in the
Astrophysics Group at Imperial College London (PPA/G/S/2000/00032).

\section*{Bibliography}

Arnaud M., Raymond J., 1992, ApJ, 398, 394 \\
Arnaud M., Rothenflug R., 1985, A\&AS, 60, 425\\
Cuntz M., Ulmschneider P., Musielak Z. E., 1998, ApJ, 493,\\
\indent L117\\
Cuntz M., Rammacher W., Ulmschneider P., Musielak Z.E., \\
\indent Saar S.H., 1999, ApJ, 522, 1053\\
Crawford H. J., Price P. B., Sullivan J. D., 1972, ApJL,\\
\indent 175, L149\\
Dere K. P., Landi E., Mason H. E., Monsignori Fossi B. C., \\
\indent Young P. R., 1997, A\&AS, 125, 149\\
Drake J. J., Smith G., 1993, ApJ, 412, 797\\
Dring A. R., Linsky J., Murthy J., Henry R. C., Moos W., \\
\indent Vidal-Madjar A., Audouze J., Landsman W., 1997,\\
\indent ApJ, 488, 760\\
Gabriel A. H., 1976, Phil. Trans. Roy. Soc. London, A291, \\
\indent 339\\
Gallagher P. T., Phillips K. J. H., Harra-Murnion L. K.,\\
\indent Keenan F. P., 1998, A\&A, 335, 733\\
Grevesse N., Sauval A. J., 1998, Space Science Reviews,\\
\indent 85, 161\\
Griffiths N. W., Jordan C., 1998, ApJ, 497, 883\\
Jordan C., 2000, Plasma Phys. Control. Fusion, 42, 415\\
Jordan C., Ayres T. R., Brown A., Linsky J. L., Simon T.,\\
\indent 1987, MNRAS, 225, 903\\
Jordan C., Brown A., 1981, in R. M. Bonnet \& A. K. \\
\indent Dupree eds., Solar Phenomena in Stars and Stellar \\
\indent Systems, p. 199, Reidel. Dordrecht, Holland, NATO\\
\indent ASIC, 68\\
Jordan C., Doschek G. A., Drake J. J., Galvin A. B., \\
\indent Raymond J. C., 1998, in ASP Conf. Ser. 154: Cool\\
\indent Stars, Stellar Systems and the Sun, Vol. 10, p. 91\\
Jordan C., McMurry A. D., Sim S. A., Arulvel M., 2001a, \\
\indent MNRAS, 322, L5\\ 
Jordan C., Sim S. A., McMurry A. D., Arulvel M., 2001b, \\
\indent MNRAS, 326, 303\\
Kelch W. L., 1978, ApJ, 222, 931\\
Kopp R. A., 1972, Solar Physics, 27, 373\\
Laming J. M., Drake J. J., Widing K. G., 1995, ApJ, 443, \\
\indent 416\\
Laming J. M., Drake J. J., Widing K. G., 1996, ApJ, 462, \\
\indent 948\\
Macpherson K. P., Jordan C., 1999, MNRAS, 308, 510\\ 
McMurry A. D., Jordan C., 2000, MNRAS, 313, 423\\
Montesinos B., Jordan C., 1993, MNRAS, 264, 900\\
Pan H. C., Jordan C., 1995, MNRAS, 272, 11\\
Philippides D., 1996, D.Phil. Thesis, University of Oxford\\
Reeves E.M., 1976, Solar Physics., 46, 53\\
Schmitt J. H. M. M., Drake J. J., Stern R. A., Haisch B. \\
\indent M., 1996, ApJ, 457, 882\\
Sim S. A., 2002, D.Phil. Thesis, University of Oxford\\
Sim S. A., Jordan C., 2003, MNRAS, 341, 517 \\
Smith G. R., Jordan C., 2002, MNRAS, 337, 666\\
Spitzer L., 1956, Physics of ionized gases, Interscience, New\\
\indent  York\\
Underwood J. H., Antiochos S. K., Vesecky J. F., 
1981,\\
\indent in R. M. Bonnet \& A. K. Dupree eds. Solar Phenomena\\
\indent in Stars and Stellar Systems, p. 227, Reidel. Dordrecht,\\
\indent Holland, NATO ASIC, 68\\
Young P. R., DelZanna G., Landi E., Dere K. P., Mason H.\\
\indent E., Landini M., 2003, A\&AS, 144, 135\\

\label{lastpage}
\end{document}